\documentclass[iop, twocolumn, letter]{aastex62}
\usepackage{amsmath}     
\usepackage{wasysym}           
\usepackage{graphicx}
\usepackage{amssymb}
\usepackage{epstopdf}
\usepackage{mathrsfs}
\usepackage{anyfontsize}
\usepackage{natbib}
\usepackage{color}
\usepackage{lipsum}
\usepackage{diagbox}

\shorttitle{Stellar Binaries Incident on SMBH Binaries}
\shortauthors{Coughlin et. al}
\begin{document}
\title{Stellar Binaries Incident on Supermassive Black Hole Binaries: Implications for Double Tidal Disruption Events, Calcium-rich Transients, and Hypervelocity Stars}
\author[0000-0003-3765-6401]{Eric R. Coughlin}
\altaffiliation{Einstein Fellow}
\affiliation{Astronomy Department and Theoretical Astrophysics Center, University of California, Berkeley, Berkeley, CA 94720}
\author{Siva Darbha}
\affiliation{Department of Physics, University of California, Berkeley, Berkeley, CA 94720}
\author{Daniel Kasen}
\affiliation{Astronomy Department and Theoretical Astrophysics Center, University of California, Berkeley, Berkeley, CA 94720}
\affiliation{Nuclear Science Division, Lawrence Berkeley National Laboratory, Berkeley, CA 94720}
\author{Eliot Quataert}
\affiliation{Astronomy Department and Theoretical Astrophysics Center, University of California, Berkeley, Berkeley, CA 94720}

\email{eric\_coughlin@berkeley.edu}

\begin{abstract}
We analyze the outcome of the interaction between a stellar binary and a supermassive black hole binary (SMBHB) by performing a large number of gravitational scattering experiments. Most of the encounters result in either the ejection of an intact binary or the ejection of two individual stars following the tidal breakup of the binary. However, tidal disruption events (TDEs) and mergers constitute a few percent of the outcomes, and double, temporally distinct TDEs (i.e., separated by at least one orbit of the supermassive black hole binary) occur at the percent level. We also demonstrate that the properties of the ejected binaries are significantly altered through the interaction with the SMBHB, and their large eccentricities {increase the merger rate} and could lead to gravitational-wave inspirals far from the nucleus of the host galaxy. We discuss our results in the context of observed tidal disruption events, hypervelocity stars, and remote supernovae, such as calcium-rich transients.
\end{abstract}

\keywords{binaries: general --- black hole physics --- galaxies: nuclei}

\section{Introduction}
\label{sec:introduction}
Discerning the presence of a supermassive black hole binary (SMBHB; \citealt{begelman80}) in the center of a galaxy is challenging (e.g., \citealt{comerford15, liu16}), and this is especially true when neither hole is active and the separation is small. One possible tool for probing such quiescent, near-merger SMBHBs is the tidal disruption of stars \citep{rees88}, as the accretion of the disrupted material onto the disrupting SMBH can be modulated on timescales comparable to the binary orbital period \citep{liu09, ricarte16, coughlin17}. Furthermore, depending on the mechanism responsible for injecting stars into the loss cone \citep{frank76, lightman77, magorrian99, stone16} and the stage of the binary inspiral, the rate of disruption can be greatly enhanced \citep{chen09, wegg11}. Thus, especially with current wide-field surveys (and the upcoming era of LSST; \citealt{ivezic08}), our inevitable detection of multiple tidal disruption events (TDEs) in the same galaxy could be highly indicative of a SMBHB at its center. 

The studies of the dynamical effects of a SMBHB on a stellar population, which ultimately lead to TDEs or stellar ejections, have largely focused on the outcome of single stars encountering the binary (e.g., \citealt{quinlan96,yu03,bromley06,sesana08,darbha18}). However, a large fraction of stars -- especially those toward the massive end -- are known to occur in binaries (e.g., \citealt{eggleton08, yuan15}). Owing to its internal degrees of freedom, a stellar binary interacting with a SMBHB can yield many more outcomes, including not just the ejection of a hypervelocity binary \citep{lu07, sesana09, guillochon15} or the disruption of one star \citep{wang18}, but the ejection of one star at the expense of capturing the other (the Hills mechanism; \citealt{hills88}), the merger of the stellar binary owing to repeated perturbations to its orbit \citep{bradnick17,liu17}, and the tidal disruption of both stars at temporally distinct times.

By performing a large number of numerical, gravitational scattering experiments between a stellar binary and a SMBHB, in this letter we attempt to gain an understanding of the relative likelihood of these outcomes. In Section \ref{sec:setup} we describe the setup of the problem and the parameters chosen for our study. Section \ref{sec:results} presents the results, and we demonstrate that two, temporally-separated TDEs can occur, albeit infrequently ($\lesssim 1\%$ of the time), from these interactions; we also discuss the properties of the ejected stellar binaries, and we give the probabilities of the various outcomes (e.g., stellar mergers, captures) as a function of stellar separation. Our conclusions are given in Section \ref{sec:conclusions}.

\section{Problem setup}
\label{sec:setup}
A stellar binary, with individual masses $m_1$ and $m_2$, total mass $m = m_1+m_2$ and semimajor axis $a_*$ (we assume here that the stellar binary and SMBHB are circular for simplicity), incident on a SMBHB, with individual masses $M_1$ and $M_2$, total mass $M = M_1+M_2$ and separation $a_\bullet$, can be described by its initial center of mass (COM) position, its COM orbit, and the orientation of the stellar semimajor axis with respect to the COM orbit. Here we will assume that the stellar binary is approaching the SMBHB from a large distance, and hence we will let the COM orbit be parabolic. We will also presuppose that the COM is in the ``pinhole'' regime, meaning that the square of its specific angular momentum, $\ell^2$, is uniformly distributed\footnote{This is a reasonable assumption if each stellar binary undergoes a number of ``collisions'' over its lifetime that randomize its COM properties and cause it to enter the loss cone from a large distance ($\gtrsim$ the SMBHB sphere of influence).}; to ensure that the stellar binary interacts strongly with the SMBHB, we will let this range be $0 \le \ell^2 \le 4GMa_\bullet$, corresponding to a stellar COM pericenter between 0 and 2$a_\bullet$. {If there are a large number of ``massive perturbers'' in the galaxy (e.g., giant molecular clouds, stellar clusters, the presence of which may be enhanced following a gas-rich merger), one can further enhance the influx of stellar binaries from the pinhole regime \citep{perets07, perets08}. Furthermore, for tight SMBHBs that are within the gravitational-wave inspiral regime, which is the scenario upon which we focus here, one expects only a handful of bound stars to be capable of diffusing into the empty loss cone within the inspiral time of the binary (\citealt{coughlin17}; we caution, however, that the merger that gave rise to the binary may also have generated a steeper cusp of stars, which could greatly enhance the rate of disruption from the empty loss cone; \citealt{stone16b}). We therefore neglect the contribution of bound stars to the disruption rate.} We further let the initial COM position be uniformly distributed over a sphere at a large distance from the binary ($\gg a_\bullet$), and the orientation of the binary be uniform over a sphere of radius $a_*$ from the position of the COM. 

The stellar binary is, in all circumstances considered here, much less massive than the SMBHB, and hence we will let the motion of the SMBHB be fixed. In this case, the four, additional variables that enter the equations are $M_1/M$, $m_1/M$, $m_2/M$, and $a_*/a_\bullet$. While the masses themselves play a role in the outcome of the encounters, we suspect that the quantity that has most influence on the survivability of the stellar binary is the ratio $a_*/a_\bullet$, as the binary tidal disruption radius scales linearly with the separation (and only as the mass ratio to the one-third power). Therefore, in our study we will let the masses be fixed at $M_1/M = 0.5$ (equal-mass SMBH binary), $m_1/M=m_2/M=0.5\times10^{-6}$ -- corresponding to Solar-like stars for $10^6M_{\odot}$ SMBHs -- and investigate the consequences of letting $a_*/a_\bullet$ vary between 0.1 and 0.0005.  

We simulated $\sim 10^5$ encounters for each choice of $a_*/a_\bullet$ using an 8th-order, explicit Runge-Kutta scheme for $\sim 1600$ binary orbits, with ejections occurring whenever a star exited a sphere of radius $50 a_\bullet$. Even though they don't enter into the equations, the tidal radius $r_{\rm t}$ must be specified in order to ``count'' disruptions. We chose $r_{\rm t}/a_\bullet = 10^{-2}$, which corresponds to $a_\bullet \simeq 1$ mpc for Solar-like stars and $10^6M_{\odot}$ SMBHs. Stellar collisions were counted when stars came within some minimum separation $r_{\rm min}$, for which we chose $r_{\rm min} = 5\times 10^{-5} a_{\bullet}$, which corresponds to $\sim 0.5 R_{\odot}$ for Solar-like progenitors. The integration was stopped if a collision occurred, while a disruption resulted in the disrupted star being removed from the simulation. We therefore do not take into account the gravitational field of the disrupted debris on the evolution of the intact star. Finally, if neither star was ejected or disrupted after $\sim 1600$ SMBHB orbits, the outcome was deemed ``inconclusive,'' which comprised $\lesssim 0.1\%$ of the outcomes.

\begin{table*}
\centering
\begin{tabular}{|c|ccccccc|}
\hline
\diagbox[width=.7in,height=0.5in,innerleftsep=-0.35in,innerrightsep=-.025in,outerrightsep=-0.11in]{{\scriptsize a$_*$/a$_\bullet$}}{\scriptsize Outcome} & {\scriptsize Binary Ejection} & {\scriptsize Double Ejection} & {\scriptsize Hills Capture} & {\scriptsize Single TDE} & {\scriptsize Prompt Double TDE} & {\scriptsize Delayed Double TDE} & {\scriptsize Merger} \\
\hline
{\scriptsize 0.0005} & {\footnotesize 92.8\%} {\footnotesize(73.8\%)} & {\footnotesize 4.11\%} {\footnotesize(4.05\%)} & {\footnotesize 0.0111\%} & {\footnotesize 0.770\%} & {\footnotesize 1.36\%} & {\footnotesize 0.0952\%} & {\footnotesize 0.838\%}   \\
{\scriptsize 0.001} & {\footnotesize87.9\%} {\footnotesize(69.1\%)} & {\footnotesize8.34\%} {\footnotesize(8.24\%)} & {\footnotesize0.00502\%} & {\footnotesize1.07\%} & {\footnotesize1.21\%} &{\footnotesize 0.132\% }&{\footnotesize 0.889\%}   \\
{\scriptsize 0.0025} & {\footnotesize76.4\%} {\footnotesize(57.8\%)} & {\footnotesize19.14\%} {\footnotesize(19.05\%)}& {\footnotesize0.0171\%} & {\footnotesize1.58\%} & {\footnotesize0.925\%} &{\footnotesize 0.153\% }&{\footnotesize 1.44\%}  \\
{\scriptsize 0.005} & {\footnotesize60.9\%} {\footnotesize(43.6\%)} & {\footnotesize34.3\%} {\footnotesize(34.2\%)} & {\footnotesize0.195\%} & {\footnotesize1.94\%} & {\footnotesize0.791\%} &{\footnotesize 0.173\% }& {\footnotesize 1.31\%}   \\
{\scriptsize 0.01} & {\footnotesize35.0\%} {\footnotesize(23.7\%)} & {\footnotesize60.5\%} {\footnotesize(58.7\%)}& {\footnotesize0.794\%} & {\footnotesize2.11\%} & {\footnotesize0.661\%} &{\footnotesize 0.186\% }& {\footnotesize0.621\%}   \\
{\scriptsize 0.05} & {\footnotesize 3.09\%} {\footnotesize(1.61\%)} & {\footnotesize92.6\%} {\footnotesize(80.8\%)} & {\footnotesize0.959\%} & {\footnotesize2.57\%} & {\footnotesize0.490\%} &{\footnotesize 0.166\% }& {\footnotesize0.0240\%}   \\
{\scriptsize 0.1} & {\footnotesize0.936\%} {\footnotesize(0.451\%)} & {\footnotesize94.6\%} {\footnotesize(82.2\%)} & {\footnotesize0.973\%} & {\footnotesize2.88\%} & {\footnotesize0.334\%} &{\footnotesize 0.179\% }& {\footnotesize0.0050\%}   \\
{\scriptsize Integrated} & {\footnotesize 45.1\%} {\footnotesize(37.7\%)} & {\footnotesize 50.6\%} {\footnotesize(49.1\%)} & {\footnotesize 0.508\%} & {\footnotesize 1.97\%} & {\footnotesize 0.661\%} & {\footnotesize 0.165\%} & {\footnotesize 0.726\%} \\
\hline
\end{tabular}
\caption{The percentage of the possible outcomes (listed in the top row) as a function of the ratio $a_*/a_\bullet$. Each number was calculated out of $\sim 10^5$ interactions between a stellar binary and the SMBHB. The last row is the time-integrated probability over the lifetime of the binary, assuming that $M_1 = M_2 = 10^6M_{\odot}$, the initial SMBHB separation is $a_\bullet = 10^4 R_{\odot}$, and $a_*$ follows a $\propto 1/a_*$ distribution. {The number in parentheses in column 1 is the percentage of ejected binaries with velocity greater than $10^3$ km s$^{-1}$ at the time of ejection (assuming the same set of SMBHB properties), while the number in parentheses in column 2 is the percentage of double ejections with {at least one} star that has a velocity greater than $10^3$ km s$^{-1}$.}} 
\label{tab:1}
\end{table*}

\begin{figure*}[htbp] 
   \centering
   \includegraphics[width=0.47\textwidth]{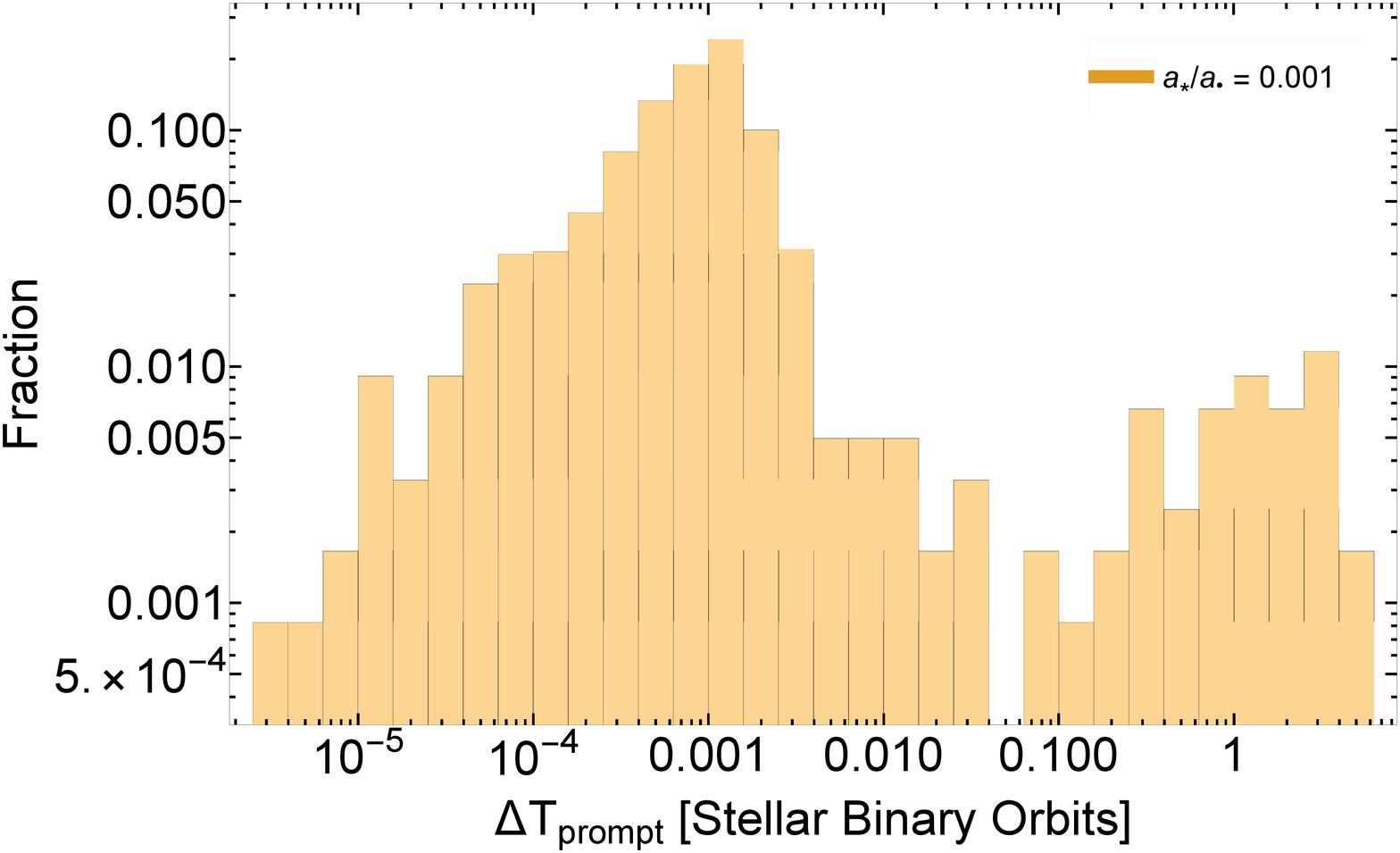} 
   \includegraphics[width=0.47\textwidth]{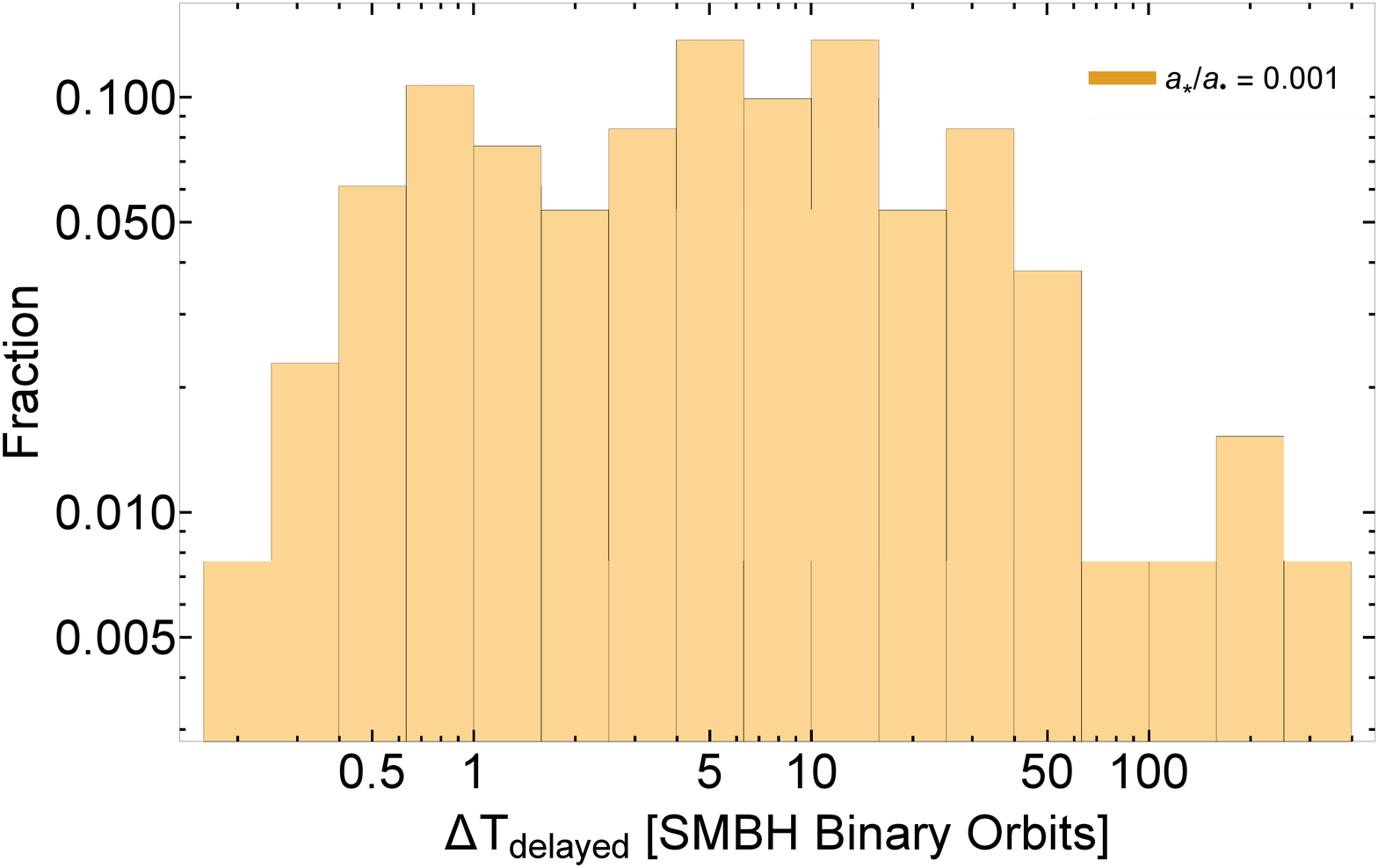} 
   \caption{Left: The distribution of time between disruptions, in units of stellar binary orbits, for prompt TDEs that occur with temporal offsets satisfying $\Delta T < 1/(2\pi)$ SMBHB orbits; Right: The distribution of time between disruptions, in units of SMBHB orbits, for delayed TDEs that occur with temporal offsets satisfying $\Delta T > 1/(2\pi)$ SMBHB orbits.}
   \label{fig:dtplot}
\end{figure*}

\section{Results}
\label{sec:results}
Table \ref{tab:1} gives the percentage of the various outcomes of the scattering experiments. {}{The last row gives the average probability of each outcome integrated over the lifetime of the SMBHB as it shrinks due to gravitational wave emission, assuming $M_1 = M_2 = 10^6M_{\odot}$, $a_\bullet = 10^4 R_{\odot}$ initially, and the PDF of the stellar binaries follows $f(a_*) \propto 1/a_*$ in the range $5 R_{\odot} \le a_* \le 10^3 R_{\odot}$\footnote{{This upper limit may be somewhat optimistic for velocity dispersions appropriate to Milky Way-type galaxies owing to their reduced survivability \citep{hills88}, but we emphasize that this specific assumption about the distribution of stellar semimajor axes \emph{only} enters into the calculation of the integrated rate; all other numbers in Table \ref{tab:1} only depend on the \emph{ratio} $a_*/a_{\bullet}$. If one reduces the upper limit to $100 R_{\odot}$, one changes the integrated percentage of ejections to $\sim 60\%$, the integrated percentage of double ejections to $\sim 35\%$, and all other integrated quantities are roughly unaltered.}} \citep{poveda07}} {(we also included a factor $\propto a_{\bullet}$ in the rate calculation to account for the SMBHB cross section, though this does not affect the results much)}. We see that the vast majority of the interactions result in either the ejection of an intact binary (binary ejection) or the ejection of both stars following the dissolution of the binary (double ejection), and the sum of these two outcomes typically totals $\gtrsim 95\%$ of the total number of events. 

{The number in parentheses in column 1 of Table 1 gives the percentage of intact binaries with an escape velocity in excess of 1000 km s$^{-1}$, while that in column 2 is the fraction of double ejections with at least one star with velocity in excess of $10^3$ km s$^{-1}$; to calculate these numbers we let the binary satisfy $M_1 = M_2 = 10^6M_{\odot}$ and $a_\bullet = 10^4 R_{\odot}$ initially. These statistics are a more direct measure of the true rate of hypervelocity ejection, as these binaries or stars would not only escape from the potential of the SMBHB, but from the galactic potential as well.} 

Hills capture -- where one star is ejected at the expense of capturing the other in a bound orbit about one of the SMBHs -- accounts for at most $\sim 1\%$ of the outcomes, and is far less likely when $a_*/a_\bullet \lesssim 0.005$. Interestingly, stellar mergers reach a peak likelihood of $\sim 1\%$ at a separation of $a_* / a_\bullet \simeq 0.0025$, which is likely due to the fact that tighter binaries are more difficult to perturb and wider binaries are more easily ripped apart. 

In addition, roughly 1-3\% of the scattering experiments result in the tidal disruption of one of the stars, and $\sim 0.5$ - 1.5\% of the encounters yield double TDEs. Of these double TDEs, a large fraction occur one after the other, or ``promptly,'' which we define to be when the temporal difference between disruptions is less than $1/(2\pi)$ of a binary orbit. As was found by \citet{mandel15}, who analyzed the deep encounter of a binary star system and a single SMBH, these prompt double TDEs occur when the pericenter of the stellar binary is within the tidal disruption radii of the stars themselves. Thus, the vast majority of these double TDEs are nearly contemporaneous. 

However, a comparable number of double TDEs occur at ``delayed'' times from one another, defined to be when the temporal offset between successive TDEs is greater than $1/(2\pi)$ of a SMBHB orbit. In these instances, the stellar binary is tidally destroyed, but \emph{both} stars are temporarily captured by the binary and eventually pass through the tidal radius of one or the other SMBH. Figure \ref{fig:dtplot} shows the distributions of the temporal offset between disruptions for the prompt disruptions (left) and the delayed disruptions (right) when $a_*/a_\bullet = 0.001$ (other values of $a_*/a_\bullet$ give very similar distributions). This figure confirms that there are two distinct classes of double disruptions: the prompt class that peaks on timescales of $\sim 0.001$ stellar binary orbits, and the delayed class that peaks on timescales of a few SMBHB orbits.

\begin{figure*}[htbp] 
   \centering
   \includegraphics[width=0.47\textwidth]{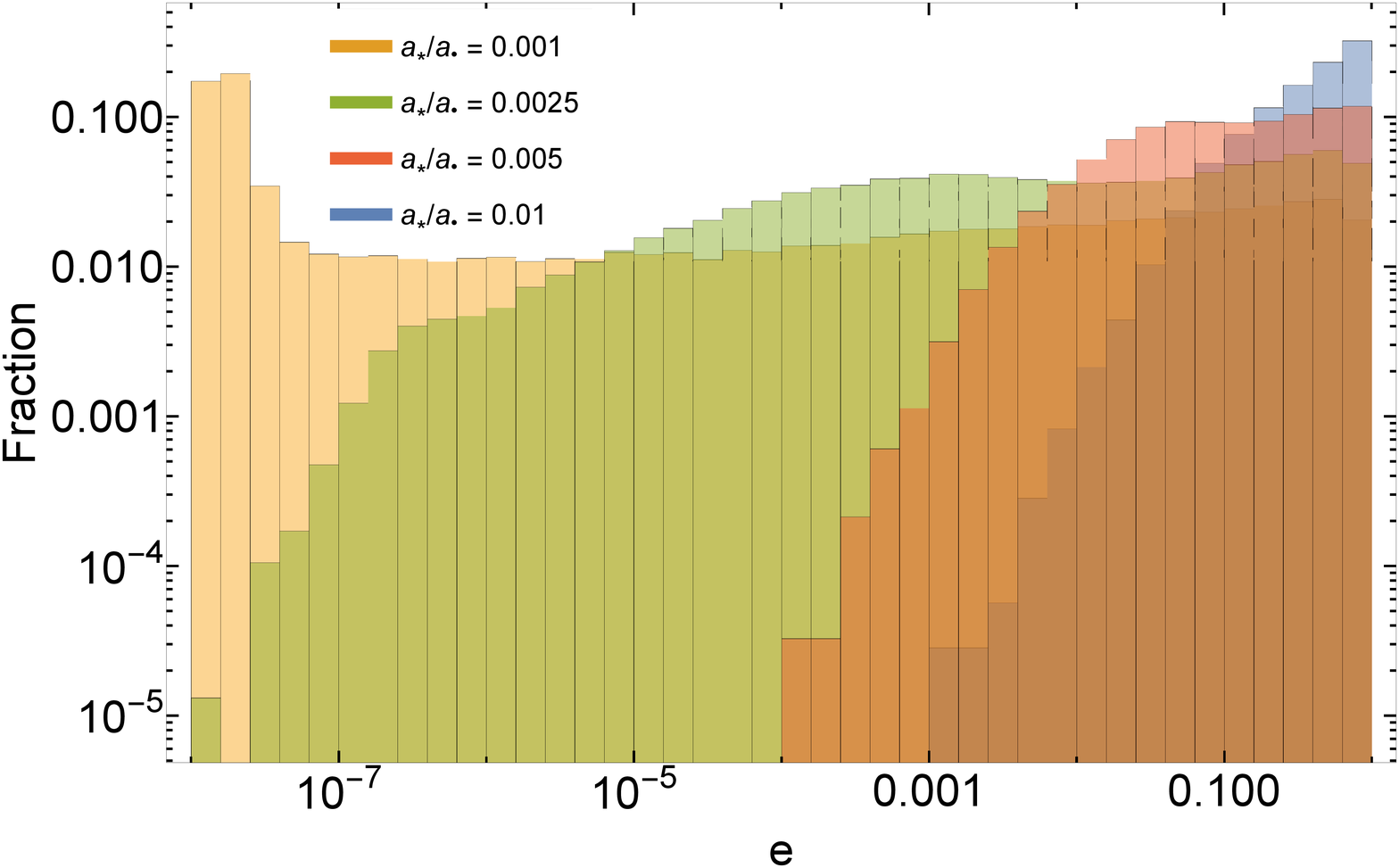} 
   \includegraphics[width=0.47\textwidth]{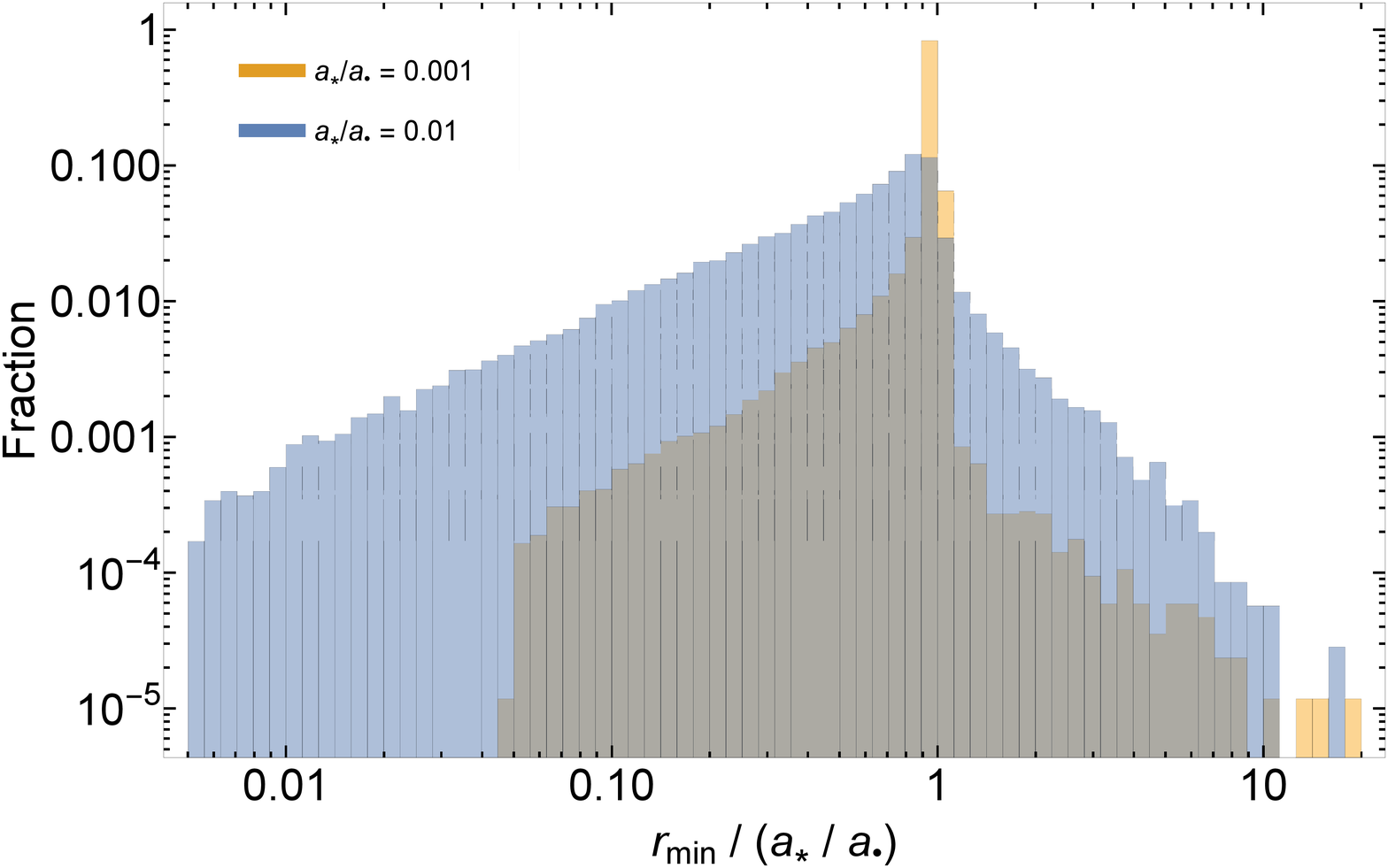} 
   \includegraphics[width=0.47\textwidth]{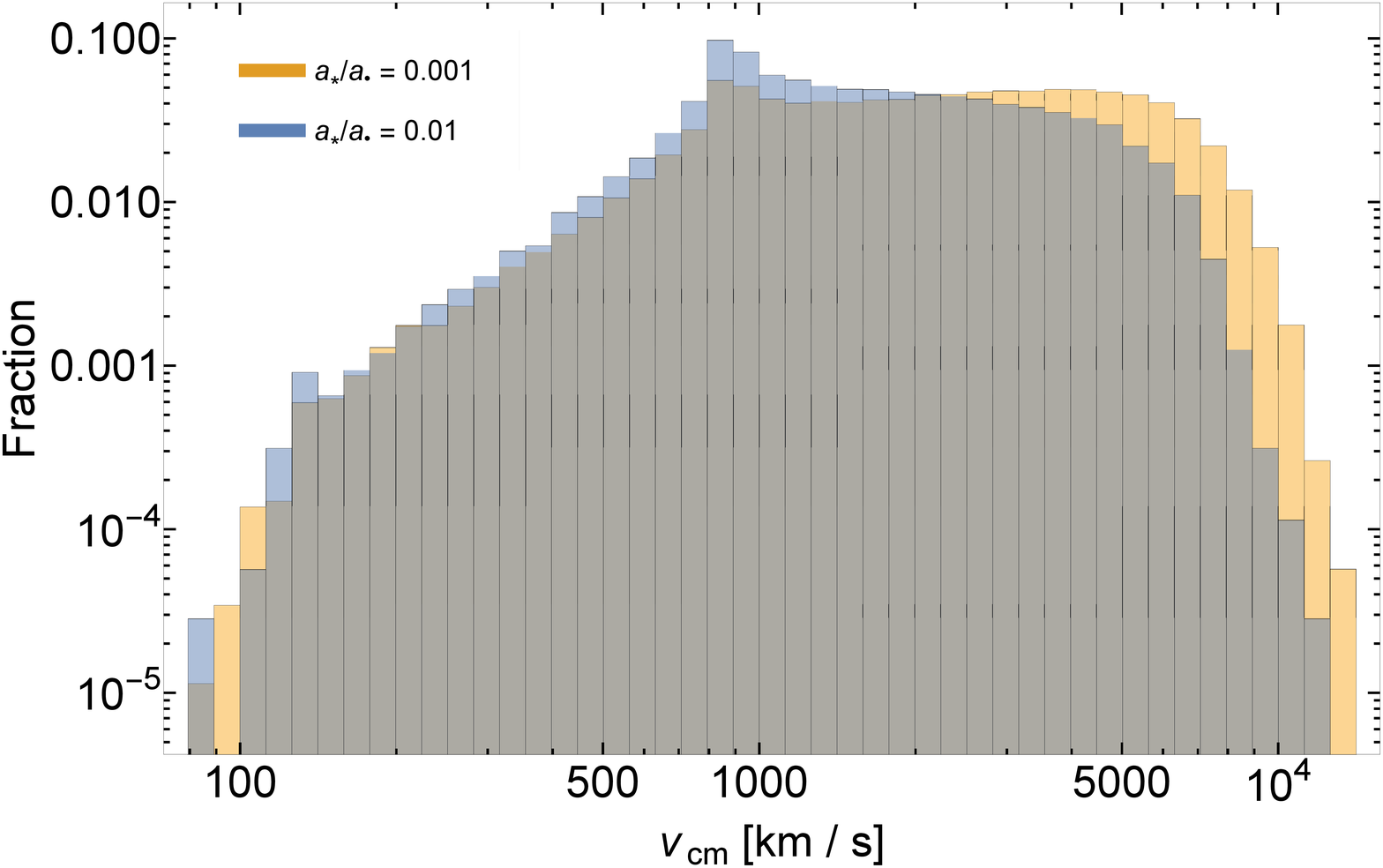} 
   \includegraphics[width=0.47\textwidth]{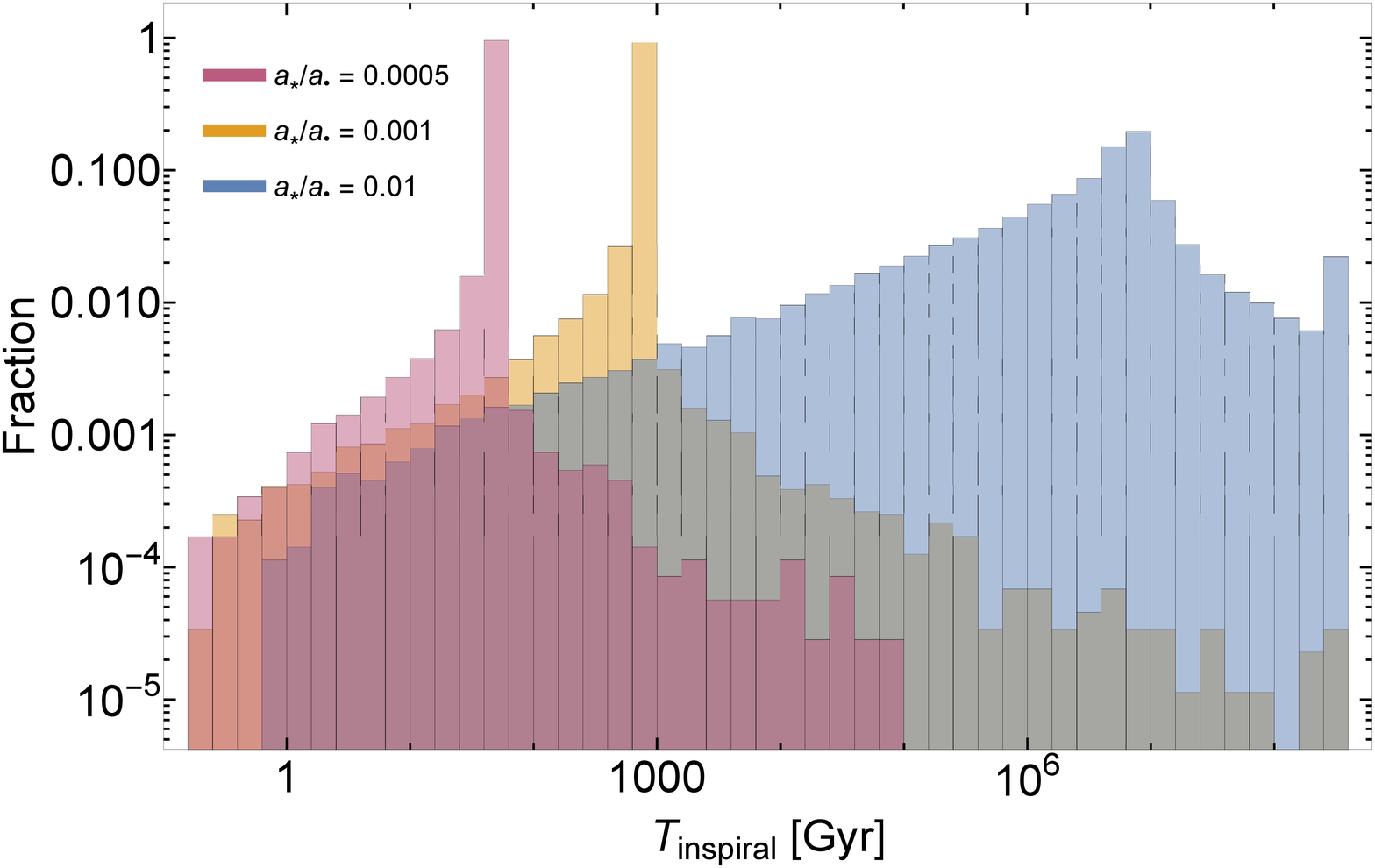}
   \caption{Top-left: The eccentricity distribution of the ejected binaries, with different colors corresponding to the ratios $a_*/a_\bullet$ indicated in the legend; Top-right: the distribution of the ejected stellar binary pericenters normalized by the initial binary separation; Bottom-left: the velocity distribution of the center of mass of the ejected binaries with $a_\bullet = 10^4 R_{\odot}$ and $M_1 = M_2 = 10^6M_{\odot}$; Bottom-right: the gravitational-wave inspiral time of the ejected binaries in years, assuming $m_1 = m_2 = 1 M_{\odot}$ and $a_{\bullet} = 10^4 R_{\odot}$.}
   \label{fig:ecc_distribution}
\end{figure*}

\begin{figure*}[htbp] 
   \centering
   \includegraphics[width=0.47\textwidth]{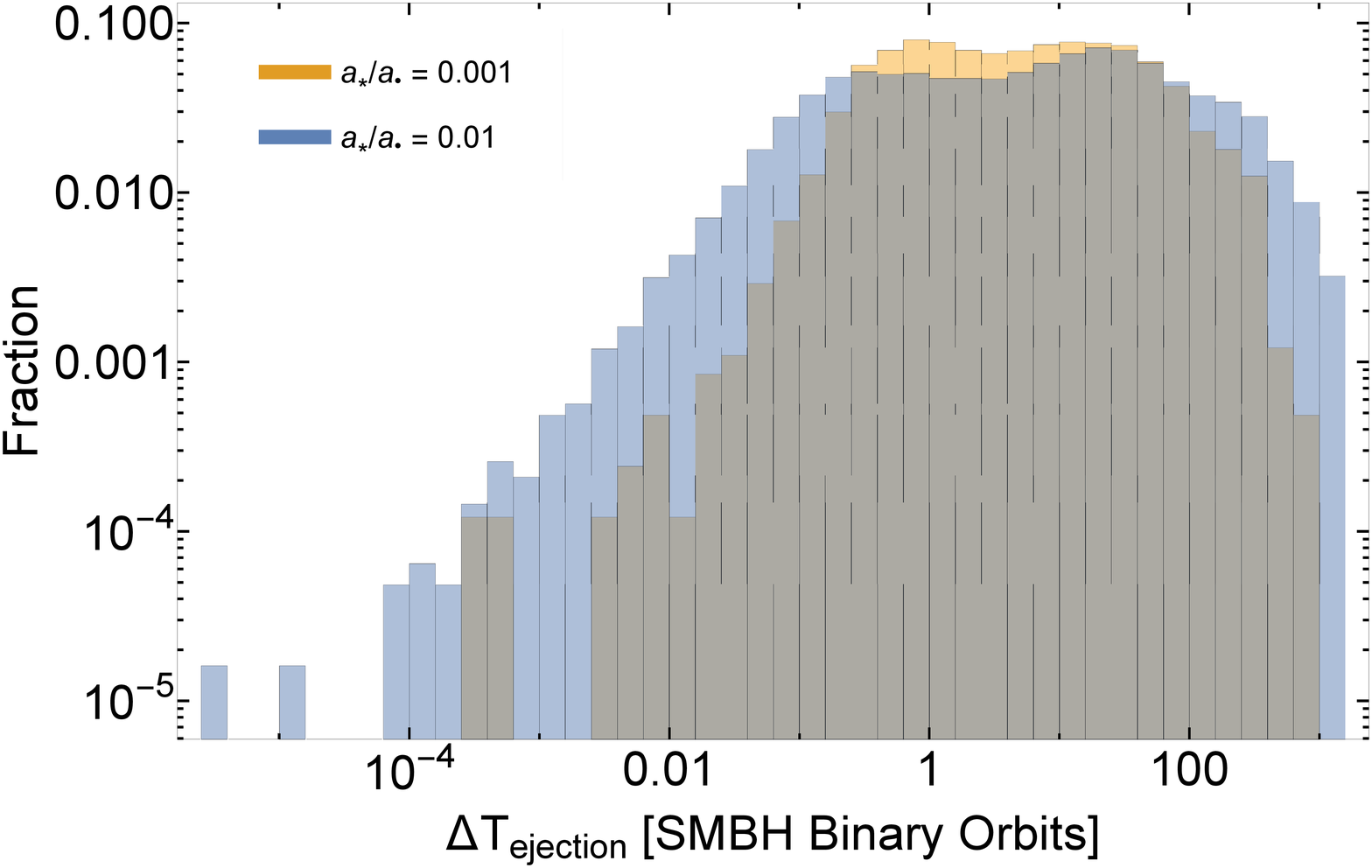} 
   \includegraphics[width=0.47\textwidth]{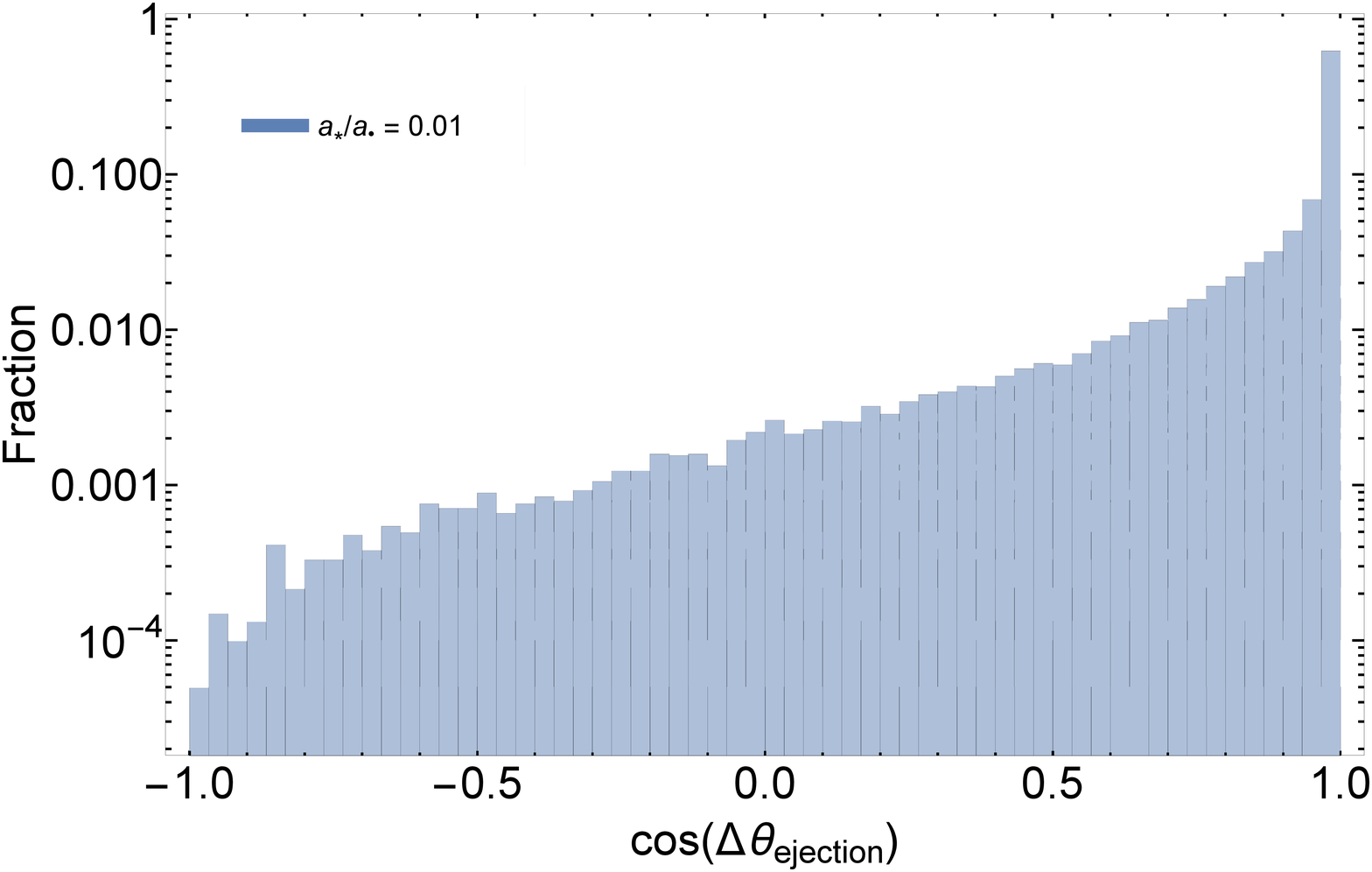} 
   \caption{Left: The temporal offset in ejections when the binary is ripped apart and both stars are subsequently ejected; Right: the cosine of the difference in polar angle between the ejected stars.}
   \label{fig:dejections}
\end{figure*}

Figure \ref{fig:ecc_distribution} gives the properties of the binaries ejected from the SMBHB, being their eccentricities (top-left), pericenters normalized by the initial semimajor axis (top-right), center of mass velocities with $a_\bullet = 10^4 R_{\odot}$ and $M_1 = M_2 = 10^6M_{\odot}$ (bottom-left), and gravitational wave inspiral times calculated from Equations 5.6 and 5.7 of \citet{peters64} with $m_1 = m_2 = 1M_{\odot}$ and $a_\bullet = 10^4 R_{\odot}$ (bottom-right). From the top-left and top-right panels we see that most binaries are perturbed from their initial states by the SMBHB, which is partially due to our requirement that the pericenter of the stellar COM be less than $2 a_{\bullet}$ from the SMBHB COM (i.e., if we had permitted larger pericenter distances in our study, then the number of unaffected binaries for $a_*/a_\bullet > 0.001$ would have been larger). It is also apparent that the distribution of stellar COM velocities is peaked around 1000 km s$^{-1}$, which is comparable to the speed of the binary $\sqrt{GM/a} \simeq 6000$ km s$^{-1}$, but the maximum attainable velocity can be more than an order of magnitude larger than this. Finally, while the inspiral distribution still peaks at a time comparable to the inspiral time of the initial binary, there are large wings induced by the relatively small number of heavily-modified orbits; for these cases, the inspiral time can be well within the age of the Universe even if that of the unperturbed orbit is not. Furthermore, this only takes into account inspirals assisted by gravitational waves, which is most relevant for compact objects; for more extended stars, the tidal dissipation timescale could be comparable or shorter (e.g., \citealt{ogilvie14}). 

Figure \ref{fig:dejections} gives the time between ejections (left panel; ejection occurs when a star reaches $50 a_\bullet$ from the SMBHB COM) and the cosine of the polar angle between ejected stars (right panel) for experiments yielding double ejections (i.e., when the binary is ripped apart and both stars are subsequently ejected). This figure demonstrates that, while there is a wide range of temporal offsets between individual stellar ejections, the angular distribution is still peaked near $\Delta\theta_{\rm ejection} = 0$ (an isotropic distribution has a flat distribution of $\cos\Delta\theta_{\rm m}$). {Also, although we did not plot the ejection angles, the ejected stars and binaries with the highest velocities stars are preferentially confined to the orbital plane of the SMBHB, which is consistent with previous findings (e.g., \citealt{sesana06}). }

{The orbit-integrated probability distribution functions for these quantities -- obtained by following an analogous weighting procedure that yielded the integrated probabilities in Table \ref{tab:1} -- appear similar to those corresponding to $a_*/a_\bullet = 0.01$. However, there are small contributions from stellar binaries with $a_*/a_\bullet < 0.01$ that widen the distributions. }

{To reduce the parameter space, here we primarily focused on circular binaries (both for the stars and SMBHs) with equal mass ratios. We did, however, assess the importance of the initial eccentricity of the stellar binary, $e_*$, by running $10^5$ encounters between a stellar binary with $e_* = 2/3$, $a_*/a_{\bullet} = 0.01$, and otherwise the same set of fiducial parameters. We found that the majority of the statistics were only slightly modified from those in Table \ref{tab:1} for $a_*/a_\bullet = 0.01$, with the one significant difference being the percentage of mergers, which increased from $\sim 0.62\%$ to $\sim 4.5\%$. Because our merger rate for circular binaries peaks at a separation of $a_*/a_\bullet \sim 0.001$, a modest increase to $\sim 1\%$ might be expected from the smaller stellar pericenter separation, being $a_*(1-e_*)/a_\bullet \simeq 0.003$. The additional increase by a factor of $\sim 4$, however, suggests that mergers are primarily driven by extreme, eccentric-Kozai-like oscillations, which increase the eccentricity to the point where the stars merge; this finding is also consistent with that of \citet{mandel15}, who found that stellar binaries approaching isolated SMBHs merged under this mechanism.}

{We also investigated the influence of the stellar mass ratio by simulating $10^5$ encounters between a circular, stellar binary with the mass of the secondary reduced by a factor of $5$ compared to the fiducial value, a stellar separation of $a_*/a_\bullet = 0.01$, and otherwise the same fiducial parameters. As was true for the eccentricity, the majority of the statistics are very similar to those in Table \ref{tab:1} with $a_*/a_\bullet = 0.01$, with the most significant differences being a reduction in the binary escape fraction to $\sim 9.3\%$ and an increase in the double escape fraction to $\sim 86\%$. These differences are likely due to the fact that the total energy of the stellar binary, $E_* = Gm_1m_2/(2a_*)$, scales in proportion to the mass ratio, and hence the binary is more easily ripped apart in this case. }

\section{Discussion and Conclusions}
\label{sec:conclusions}
Owing to the chaotic, gravitational interactions between the stars in a stellar binary and the black holes in a SMBHB, we have shown that scattering events between the two can generate a variety of dynamical outcomes. In particular, while ejections of an intact stellar binary and individual stellar ejections following the tidal separation of the binary dominate the statistics, more exotic end states -- including stellar mergers, single and double tidal disruptions -- can occur, albeit with reduced likelihoods. Table \ref{tab:1} summarizes the relative probabilities of these occurrences as a function of $a_*/a_\bullet$, the ratio of the stellar binary semimajor axis to that of the SMBHB.

In addition to double TDEs that occur when the stellar COM pericenter comes within the tidal disruption radius of the individual stars, which result in ``prompt,'' or nearly-contemporaneous disruptions by one SMBH \citep{mandel15}, we also found that a comparable fraction of double TDEs occur after the stellar binary is dissociated and with large temporal offsets ($\gtrsim 1/(2\pi)$ SMBHB orbits). From Table \ref{tab:1}, we see that the fraction of delayed, double TDEs is approximately independent of the ratio $a_*/a_\bullet$, which likely results from a competition between tidal stripping (easier for wider binaries) and capture (harder for wider binaries). Figure \ref{fig:dtplot} shows the distribution of the time between disruptions, indicating that the delay can be between a fraction and hundreds of SMBHB orbits. Two such delayed TDEs resulting from the disruption of a binary could explain the extremely luminous, double-peaked transient ASASSN-15lh \citep{dong16,brown16,godoy-rivera16}; while a TDE interpretation has already been investigated for this event, both in the single-star-SMBH \citep{leloudas16} and single-star-SMBHB \citep{coughlin18} scenarios (an extremely energetic and exotic supernova could have also powered the emission; \citealt{chatzopoulos16}), two, temporally-offset TDEs following the separation of a stellar binary naturally explains the rebrightening {(the lack of hydrogen and helium emission lines is an additional, puzzling aspect of this event, but may tentatively be explained by optical depth effects; \citealt{roth16})}. 

\begin{figure*}[htbp] 
   \centering
   \includegraphics[width=0.47\textwidth]{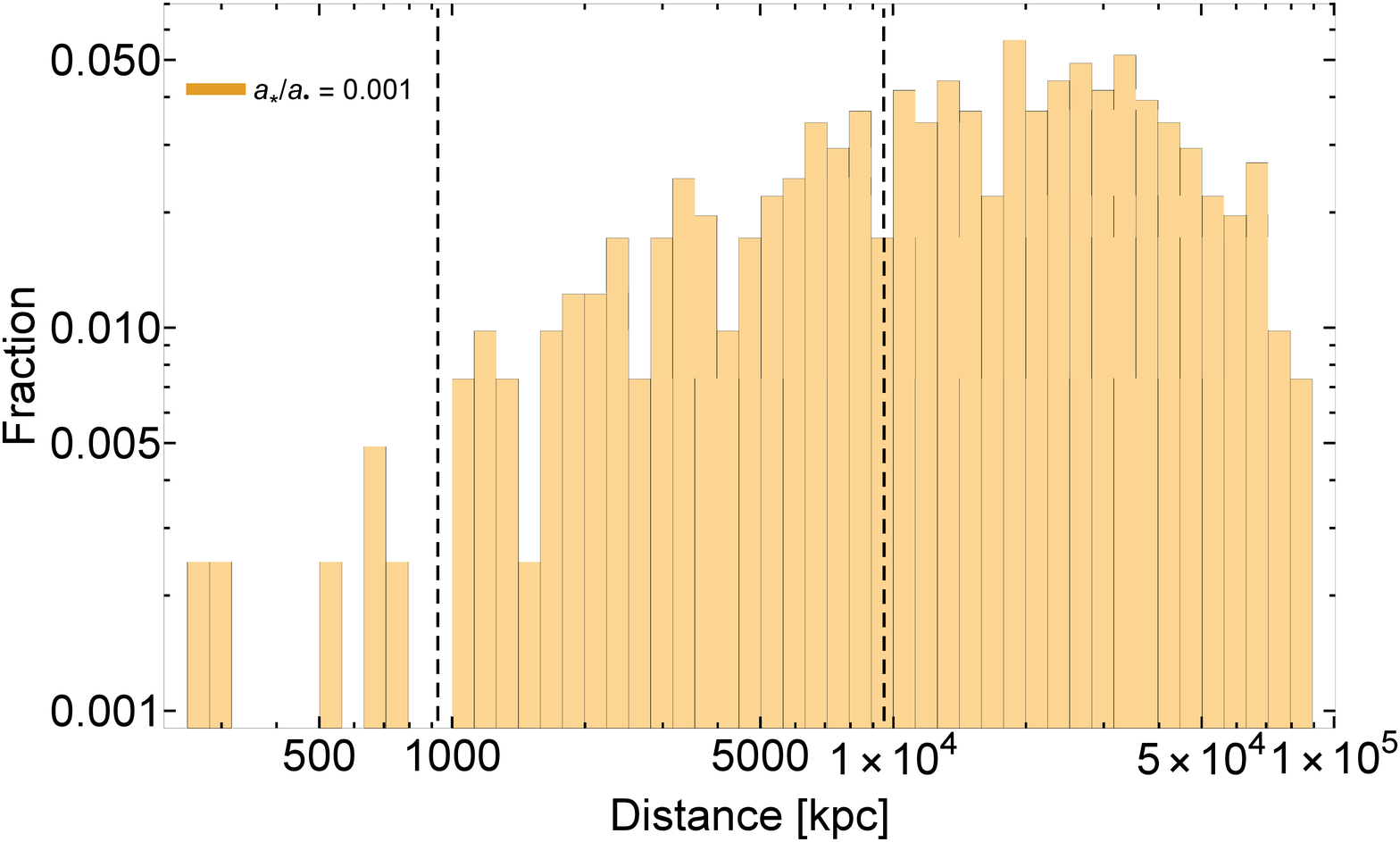} 
   \includegraphics[width=0.47\textwidth]{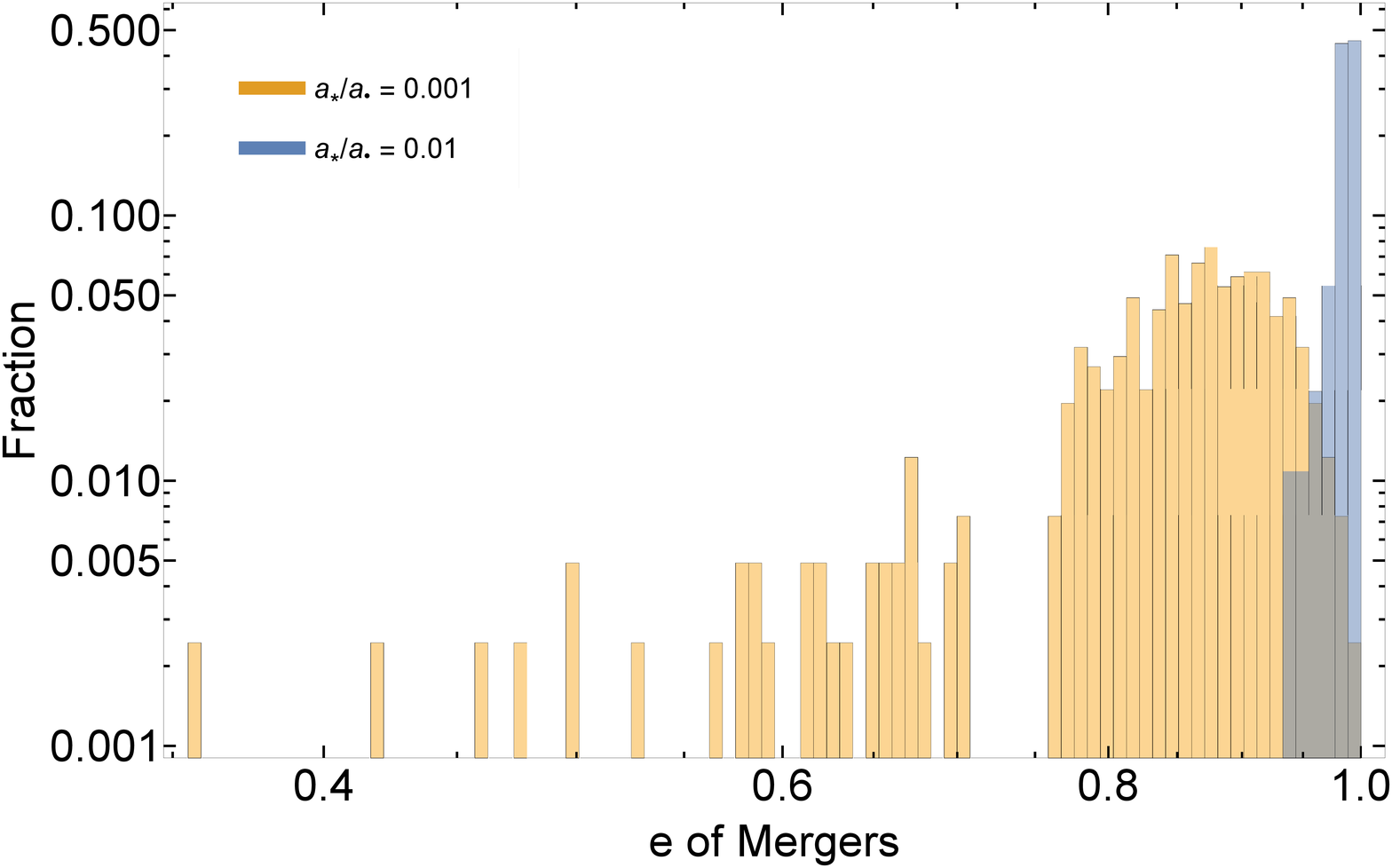}
   \caption{Left: The distribution of the distance, in kpc, of the ejected stellar binaries at the time of gravitational-wave coalescence when $a_*/a_\bullet = 0.001$ and $a_\bullet = 10^4 R_{\odot}$; here we only counted the binaries with an inspiral time less than 10 Gyr, and we did not account for the deceleration caused by the gravitational potential of the host galaxy {(we also note that the ejection velocity scales as $a_{\bullet}^{-1/2}$, so wider binaries -- which are more likely -- and the same physical range of $a_*$ would reduce the distances considerably)}. The vertical, dashed lines give, from left to right, the maximum distance attained by the distribution if we restrict the possible inspiral times to 100 Myr and 1Gyr. Right: The eccentricities (at the time of ejection) of binaries that undergo mergers in less than 10 Gyr.}
   \label{fig:distance}
\end{figure*}

Ejected, intact binaries constitute the vast majority of the outcomes for $a_*/a_\bullet \lesssim 0.01$. Investigating Figure \ref{fig:ecc_distribution}, a fraction ($\sim$ few $\times$ 0.1 -- 1\%) of ejected binaries merge in significantly less time than the original gravitational-wave inspiral time of the binary and within the age of the Universe. Also, given the relatively large velocities imparted to the binaries, their nuclear separation from the host galaxy at the time of merger is substantial; the left-hand panel of Figure \ref{fig:distance} shows the distribution of distances from the nucleus, calculated by taking the product of the ejected velocity and the inspiral time, for $a_*/a_\bullet = 0.001$, $a_\bullet = 10^4 R_{\odot}$, and $M_\bullet = 10^6M_{\odot}$ if we demand that the gravitational-wave merger time be less than 10 Gyr. The vertical, dashed lines at $10^4$ ($10^3$) shows the maximum achievable distance if we restrict the inspiral time to less than 1 (.1) Gyr. 

{}{It has recently been suggested \citep{foley15} that Calcium-rich transients \citep{filippenko03, perets10, kasliwal12, lyman14} could be the product of gravitational-wave inspirals of white dwarf (WD)-white dwarf binaries, their large galactic offsets caused by the ejection of the stellar binary following its interaction with a SMBHB. Our inferred distances at the time of inspiral, while likely overestimates of the true distances owing to our small SMBHB separation {(the ejected velocity scales as $\propto a_\bullet^{-1/2}$, so the same range of $a_*$ and larger $a_\bullet$ would significantly reduce the distances)} and our neglect of the galactic potential and tidal dissipation within the stellar binary itself, confirm that this aspect of Ca-rich supernovae can be reproduced with this mechanism. }

{}{The theoretical rate at which these inspirals take place is uncertain from our analysis alone, as we only explored a restricted range of parameter space and we are not accounting for a number of priors. However, if the binary is already composed of WDs when it reaches the SMBHB, then the rate of inspirals is $\sim 10^{-4} \times a_\bullet/a_{\rm t}\times F_{\rm SMBHB}\times F_{\rm merger} \times F_{\rm WDWD} \sim 10^{-4} - 10^{-5}$ gal$^{-1}$ yr$^{-1}$, where {$F_{\rm SMBHB} \le 1$ is the fraction of SMBHs in binaries}, $F_{\rm merger} \sim 10^{-1} - 10^{-2}$ is the fraction of ejected systems that merge in a Hubble time and $F_{\rm WDWD} \lesssim 10^{-1}$ \citep{brown11} is the fraction of WD-WD binaries encountering the SMBHB.\footnote{{Since one of the WDs must be Helium-rich, however, this number is likely reduced by at least another order of magnitude.}} If the binary is still in the main sequence phase at the time of ejection, then the total number of ejected binaries is likely higher, but those that survive intact through the common envelope phase is uncertain. Regardless, our findings indicate that the rate of mergers from such systems is probably only $10^{-1} -10^{-3}$ the type-Ia rate of $0.001 - 0.01$ gal$^{-1}$ yr$^{-1}$ (e.g., \citealt{scannapieco05,li11,maoz12}), which is toward the low end of the estimated rate of Calcium-rich transients (e.g., \citealt{perets10, kasliwal12, foley15}; more recent work, however, indicates that the rate could be as high as the type-Ia rate; Frohmaier et al.~in prep.). We note, however, that the theoretical and observational rates could be brought closer to agreement if wider SMBHBs boost the rate geometrically, Calcium-rich transients repeat (which could occur, e.g., if a phase of mass transfer ignites a more mild explosion near the pericenter of the orbit), or the rate of injection of stars into the SMBHB loss cone is intrinsically higher.}

{Potentially-discrepant rates aside, there are other aspects of Calcium-rich transients that seem to defy the explanation that they exclusively originate from WD-WD binaries kicked from the centers of galaxies. For example, the event iPTF15eqv displayed a mixture of properties appropriate to Calcium-rich transients and core-collapse supernovae \citep{milisavljevic17}. This source also lacked significantly Doppler-shifted lines, a feature that must necessarily be present in the spectrum if the large velocity imparted to the binary from the SMBHB (see Figure \ref{fig:ecc_distribution}) explains the nuclear offset of the source (though there could be line of sight effects that complicate this trend). Furthermore, Ca-rich transients seem to trace an older population of stars, and type-Ia supernovae from such an older population can be located in the outskirts of galaxies and are therefore consistent with the lack of any inferred kick velocity \citep{perets14}. There is therefore evidence -- not only from our lower-than-observed rate calculations here -- that not all Calcium-rich transients arise from this avenue of WD-WD binary ejection.}

{}{The binaries that undergo mergers within a Hubble time have very high eccentricities at the time of ejection -- especially those with larger initial separations -- as shown in the right-hand panel of Figure \ref{fig:distance}. In these highly eccentric orbits, resonances between the orbital timescale and the stellar Eigenmodes could be strong enough to tidally detonate the stars \emph{before} the gravitational-wave inspiral time (e.g., \citealt{rathore05,fuller11,fuller12,burkart13}). Finally, if the system is initially composed of two main sequence stars that then evolve through a phase of common envelope evolution, the passage of the compact object through the envelope of the companion on such a highly-eccentric orbit could produce another avenue of reducing the separation more rapidly. Although the outcome and observational appearance of tidally-induced mergers are unclear, this should be a source of some variety of transient well-separated from any host.}

Finally, the ejection of two stars following the disruption of the binary (double ejections) has interesting implications for the detection of hypervelocity stars (e.g., \citealt{brown15}). From Figure \ref{fig:dejections}, we see that $\gtrsim 60\%$ of double ejections yield angular separations $\gtrsim 0.99\text{ rad} \simeq 8^{\circ}$, corresponding to $\sim $ few kpc-scale separations between the stars once they recede to $\sim 100$ kpc from the galaxy. If the two ejected stars formed from the same protostellar cloud, and hence that the original stellar binary was preserved from the site of star formation, two such hypervelocity stars could look very similar spectroscopically but would have spatial offsets on the order of kpc. The detection of such spatially-distinct, spectroscopically-similar, hypervelocity stars would then be a strong indication of the presence of a SMBHB in the center of the host galaxy.

\acknowledgements
ERC acknowledges support from NASA through the Einstein Fellowship Program, grant PF6-170150. This work was supported in part by a Simons Investigator award from the Simons Foundation (EQ) and the Gordon and Betty Moore Foundation through Grant GBMF5076. This work was supported by the National Science Foundation under Grant No.~1616754. {We thank the referee for constructive comments and suggestions.}

\bibliographystyle{aasjournal}

\end{document}